\documentclass[aps,prd,onecolumn,groupedaddress,showpacs,nofootinbib,amssymb
]{revtex4}
\usepackage[dvips]{graphicx,color}
\usepackage{amssymb}
\usepackage{amsmath}
\usepackage{graphicx}
\usepackage{amsfonts}
\usepackage{bm}
\newcommand{\be}{\begin{equation}}
\newcommand{\ee}{\end{equation}}
\newcommand{\bea}{\begin{eqnarray}}
\newcommand{\eea}{\end{eqnarray}}
\newcommand{\beaa}{\begin{eqnarray*}}
\newcommand{\eeaa}{\end{eqnarray*}}

\newcommand{\nn}{\nonumber \\}
\newcommand{\e}{\mathrm{e}}
\newcommand{\tr}{\mathrm{tr}\,}
\newcommand{\sign}{\mathrm{sign}\, }

\allowdisplaybreaks[4]

\begin{document}

\title{ Singular Accelerated Evolution in massive $F(R)$ bigravity}
\author{S.~Nojiri,$^{1,2}$\,\thanks{nojiri@gravity.phys.nagoya-u.ac.jp}
S.~D.~Odintsov,$^{3,4}$\,\thanks{odintsov@ieec.uab.es}
V.~K.~Oikonomou,$^{5,6}$\,\thanks{v.k.oikonomou1979@gmail.com}}
\affiliation{$^{1)}$ Department of Physics, Nagoya University, Nagoya 464-8602, 
Japan \\
$^{2)}$ Kobayashi-Maskawa Institute for the Origin of Particles and the
Universe, Nagoya University, Nagoya 464-8602, Japan \\
$^{3)}$Institut de Ciencies de lEspai (IEEC-CSIC),
Campus UAB, Carrer de Can Magrans, s/n\\
08193 Cerdanyola del Valles, Barcelona, Spain\\
$^{4)}$ ICREA, Passeig LluA­s Companys, 23,
08010 Barcelona, Spain\\
$^{5)}$ Tomsk State Pedagogical University, 634061 Tomsk, Russia\\
$^{6)}$ Lab. Theor. Cosmology, Tomsk State University of Control Systems
and Radioelectronics, 634050 Tomsk, Russia (TUSUR)\\
}

\begin{abstract}
The possibility to have singular accelerated evolution in the context of 
$F(R)$ bimetric gravity is investigated. Particularly, we study two 
singular models of 
cosmological evolution, one of which is a singular modified version of the 
Starobinsky $R^2$ inflation model. As we demonstrate, for both models in some 
cases, the slow-roll parameters become singular at the Type IV singularity, a 
fact that we interpret as a dynamical instability of the theory under study. 
This dynamically instability may be an indicator of graceful exit from 
inflation and we thoroughly discuss this scenario and the interpretation of the 
singular slow-roll parameters. Furthermore, it is demonstrated that for 
some versions of $F(R)$ bigravity, singular inflation is realized in 
consistent way so that inflationary indices are compatible with Planck 
data. Moreover, we study the late-time behavior of the 
two singular models and we show that the unified description of early and 
late-time acceleration can be achieved in the context of bimetric $F(R)$ 
gravity.
\end{abstract}


\maketitle

\section{Introduction}

Since the striking discovery of late-time acceleration in the late 90's 
\cite{riess}, many theoretical descriptions have been proposed to model this 
late-time acceleration. This late-time acceleration contributes at almost 76$\% 
$  to the total energy density of the Universe, and it is known as dark energy. 
Modified theories of gravity provide a successful and self-consistent 
description of dark energy (see reviews \cite{reviews1}).
On the same time, modified gravity may also successfully describe the 
early-time acceleration, i.e. inflationary stage, giving the possibility 
for unified description of inflation and dark energy \cite{NO1}.
There also exist more traditional mechanisms that can successfully 
describe the dark energy as kind of effective unusual fluid (see for 
example reviews \cite{capo,capo1,peebles}) without need to 
modify the gravitational theory. One of the ways that 
may provide a promising description of the late-time acceleration (as 
eventually of the inflation too) within modified gravity paradigm, is to 
give a mass 
to the spin-2 particle, the graviton, i.e. to consider massive gravity.

However, giving a mass to the graviton is not a trivial task, with the first 
attempts towards this goal dated back in the 40's, when Fierz and Pauli 
\cite{fierzpauli} attempted to describe in a linear way a massive graviton 
theory. Their linear theory gave rise to a discontinuity in the observable 
physical quantities, which discontinuity can be avoided in the context of a 
non-linear massive gravity theory. The non-linearities however always bring 
along a negative norm ghost field, which is known as the Boulware-Deser ghost, 
which eventually renders the theory unstable. Thus the massive gravity concept 
remained intangible until recently, where the interest in these massive gravity 
theories was renewed, since as was demonstrated in 
Refs.~\cite{bigravfun1,bigravfun2,bigrahasan}, it is possible to have a 
non-linear massive gravity, free of the Boulware-Deser ghost, at the decoupling 
limit \cite{bigravfun2} and in the full theory \cite{bigrahasan}. For some 
recent works on massive bigravity theories, see 
\cite{bigra1,bigra2,bigra3,bigra4,bigra5,bigra6,viser}.

In the context of bimetric
massive gravity, there are two metrics, one describing the physical Universe we 
observe and one background fiducial metric, and there exist solutions in which 
the physical metric is the Friedmann-Robertson-Walker (FRW) metric and the 
background metric is Minkowski metric, so non-flat, and also there exist 
solutions for which the background metric is dynamical and non-flat, for 
example a  FRW one, and the physical metric is also a FRW one \cite{viser}. In 
principle, the massive gravity theory has many alternative descriptions, with 
all the descriptions however agreeing on the fact that two metrics are 
necessary to exist, say $g^{\mu \nu}$ and $f^{\mu \nu}$, and the ``massive'' 
interaction term being a scalar function of $g^{\mu \nu}f_{\mu \nu}$.

Although the first attempts in massive bigravity theory made use of a flat and 
non-dynamical background metric $f^{\mu \nu}$, since if the fiducial metric is 
chosen to be a FRW one, new non-linear ghost instabilities appear 
\cite{bigra5}. However, it has been shown that \cite{bigrahasan,bigra1,viser} 
in the bimetric theories of gravity, it is possible to have a non-flat and 
dynamical fiducial metric, and in this case the theory is free of ghosts, 
again. Actually, as was demonstrated in \cite{viser}, solutions of bimetric 
gravity in the limit where the kinetic term of the background fiducial metric 
vanishes, there exist massive gravity solutions compatible with a dynamical and 
non-flat metric. In addition, the extension of bigravity theories for the 
$F(R)$ gravity case has been performed in Refs.~\cite{bigra6}.

In this paper the focus is on providing a description of singular cosmological 
dynamics in the context of $F(R)$ bimetric gravity theory. The singularity that 
we shall take into account is the Type IV singularity, which is a finite-time 
timelike singularity \cite{Nojiri:2005sx}. According to the classification of 
finite-time singularities \cite{Nojiri:2005sx}, the Type IV singularity is the 
most ``harmless'' one, since the Universe may smoothly pass through it without 
having catastrophic consequences for the observable quantities that can be 
defined on the spacelike three dimensional hypersurface defined at the time 
instance that the singularity occurs. Therefore, unlike crushing type 
singularities, firstly studied in a concrete way by Hawking and Penrose 
\cite{hawkingpenrose}, like for example the Big-Rip singularity \cite{ref5}, 
the Type IV singularity does not affect the observables catastrophically, but 
affects strongly the dynamics and the slow-roll expansion of inflationary 
evolution \cite{inflation}, as was shown in \cite{noo4}. Actually, the graceful 
exit procedure may be enhanced by the presence of a Type IV singularity, as was 
shown in \cite{noo4}. For recent work on the Type IV singularities see 
\cite{noo4,Barrow:2015ora,noo1,noo2,noo3}, while for important earlier 
works on sudden singularities see \cite{Barrow:2004hk,barrow}, and
  for alternative works on the graceful exit issue, see \cite{lehners}. 
Since the inflationary era \cite{inflation} is very important for present time 
observations, we shall investigate how a Type IV singularity can affect the 
dynamics of this era, in the context of bimetric gravity. Particularly, after 
presenting in brief the general bimetric gravity model we shall work on, we 
shall assume a quite general and simple cosmological evolution, developing a 
Type IV singularity, and we investigate which bimetric gravity model can 
successfully generate such an evolution. For the resulting model, we shall 
calculate the slow-roll parameters \cite{barrowslowroll}, and as we 
demonstrate, in some cases these become strongly divergent at the singularity 
point. We discuss the implications of these divergences in a later section 
where we claim that these indicate instabilities in the dynamical evolution, 
which in turn show that the cosmological solution which described the evolution 
up to the moment the singularities occur, ceases to be the final attractor of 
the theory, thus graceful exit from inflation may occur at the Type IV 
singularity. Furthermore, we study a Type IV singular version of the $R^2$ 
Starobinsky model \cite{starobinsky}, which as we show, also results in 
singular slow-roll parameters. Finally, we study the late-time behavior of our 
cosmological solutions by studying the corresponding equation of state, and in 
addition we propose some generalized Type IV singular cosmological evolutions, 
which can describe early and late-time acceleration, but also the matter 
domination era, by using only one model.

This paper is organized as follows: In section II we describe the theoretical 
framework of the $F(R)$ bimetric gravity we shall use and in section III and IV 
after providing the essential information for the finite-time singularities, we 
investigate which bimetric gravity theories can generate the singular 
cosmological evolutions we mentioned earlier. We also calculate the 
corresponding slow-roll parameters and also we investigate when these become 
singular at the time instance that the singularity occurs. In the end of 
section IV, we discuss the implications of the singularities appearing in the 
slow-roll parameters and consequently the implications on the inflationary 
dynamics. The late-time behavior and the unification of early and late-time 
dynamics is presented in section V, while the conclusions appear in the end of 
the paper.

\section{Bigravity Essentials}

In this section we briefly review the theoretical framework of $F(R)$ bigravity 
and also the relevant formalism. For more details on this we refer to 
Ref.~\cite{bigra6}. The general Jordan frame action of 
$F(R)$ 
bigravity contains two auxiliary fields $\varphi$ and $\xi$, and is given 
below,
\begin{align}
\label{FF1}
S_{F} =& M_f^2\int d^4x\sqrt{-\det f^{\mathrm{J}}}\,
\left\{ \e^{-\xi} R^{\mathrm{J}(f)} +  \e^{-2\xi} U(\xi) \right\} \nn
& +2m^2 M_\mathrm{eff}^2 \int d^4x\sqrt{-\det g^{\mathrm{J}}}\sum_{n=0}^{4}
\beta_n
\e^{\left(\frac{n}{2} -2 \right)\varphi - \frac{n}{2}\xi} e_n
\left(\sqrt{{g^{\mathrm{J}}}^{-1} f^{\mathrm{J}}} \right) \nn
& + M_g^2 \int d^4 x \sqrt{-\det g^{\mathrm{J}}}
\left\{ \e^{-\varphi} R^{\mathrm{J}(g)} +  \e^{-2\varphi} V(\varphi) \right\}
+ \int d^4 x \mathcal{L}_\mathrm{matter}
\left( g^{\mathrm{J}}_{\mu\nu}, \Phi_i \right)\, .
\end{align}
In Eq.~(\ref{FF1}), $R^{(g)}$ and $R^{(f)}$ represent the scalar curvatures for 
the Jordan frame metrics $g^{\mathrm{J}}_{\mu\nu}$ and 
$f^{\mathrm{J}}_{\mu\nu}$, respectively. Also in the theory there exist two 
mass scales, the two Planck masses $M_f$ and $M_g$, and we define 
$M_\mathrm{eff}$ to be equal to,
\be
\label{Meff}
\frac{1}{M_\mathrm{eff}^2} = \frac{1}{M_g^2} + \frac{1}{M_f^2}\, .
\ee
In addition, we also define the tensor $\sqrt{g^{-1} f}$ by using the square 
root of ${g^{\mathrm{J}}}^{\mu\rho} f^{\mathrm{J}}_{\rho\nu}$, so that the 
following holds true,
$\left(\sqrt{g^{-1} f}\right)^\mu_{\ \rho} \left(\sqrt{g^{-1}
f}\right)^\rho_{\ \nu} = g^{\mu\rho} f_{\rho\nu}$.
The symbols $e_n(X)$'s are defined for a general tensor $X^\mu_{\ \nu}$ in the 
following way,
\begin{align}
\label{ek}
& e_0(X)= 1  \, , \quad
e_1(X)= [X]  \, , \quad
e_2(X)= \tfrac{1}{2}([X]^2-[X^2])\, ,\nn
& e_3(X)= \tfrac{1}{6}([X]^3-3[X][X^2]+2[X^3])
\, ,\nn
& e_4(X) =\tfrac{1}{24}([X]^4-6[X]^2[X^2]+3[X^2]^2
+8[X][X^3]-6[X^4])\, ,\nn
& e_k(X) = 0 ~~\mbox{for}~ k>4 \, ,
\end{align}
where the trace of the tensor $X^\mu_{\ \nu}$: $[X]=X^\mu_{\ \mu}$ is denoted 
by $[X]$ in all the above equations and in the equations to follow, when used. 
By varying the action (\ref{FF1}), with respect to $\varphi$ and $\xi$, we can 
obtain algebraic equations that relate the Ricci scalars to the auxiliary 
fields $\varphi$ and $\xi$. The resulting algebraic equations can, in 
principle, be solved algebraically with respect to the auxiliary fields 
$\varphi$ and $\xi$, and upon substituting the resulting expressions into 
(\ref{FF1}) we can obtain the $F(R)$ bigravity action which does not include 
the auxiliary scalars $\varphi$ and $\xi$.

By conformally transforming the Jordan frame metric tensors 
$g^{\mathrm{J}}_{\mu\nu}$ and $f^{\mathrm{J}}_{\mu\nu}$ in the following way,
\begin{equation}
\label{conftransform}
g_{\mu\nu} \to \e^{-\varphi} g^{\mathrm{J}}_{\mu\nu}\, ,\quad f_{\mu\nu}\to 
\e^{\xi} f^{\mathrm{J}}_{\mu\nu}\, ,
\end{equation}
the action of Eq.~(\ref{FF1}) can be transformed as follows,
\begin{align}
\label{total}
S_{F} =& S_\mathrm{bi} + S_\varphi + S_\xi\, , \\
\label{bimetric}
S_\mathrm{bi} =&M_g^2\int d^4x\sqrt{-\det g}\,R^{(g)}+M_f^2\int d^4x
\sqrt{-\det f}\,R^{(f)} \nonumber \\
&+2m^2 M_\mathrm{eff}^2 \int d^4x\sqrt{-\det g}\sum_{n=0}^{4} \beta_n\,
e_n \left(\sqrt{g^{-1} f} \right) \, , \\
\label{Fbi1}
S_\varphi =& - M_g^2 \int d^4 x \sqrt{-\det g}
\left\{ \frac{3}{2} g^{\mu\nu} \partial_\mu \varphi \partial_\nu \varphi
+ V(\varphi) \right\} + \int d^4 x \mathcal{L}_\mathrm{matter}
\left( \e^{\varphi} g_{\mu\nu}, \Phi_i \right)\, ,\\
\label{Fbi7b}
S_\xi =& - M_f^2 \int d^4 x \sqrt{-\det f}
\left\{ \frac{3}{2} f^{\mu\nu} \partial_\mu \xi \partial_\nu \xi
+ U(\xi) \right\} \, .
\end{align}
In the present paper we shall consider the simplest case, in which no matter 
fluids are present, so the Einstein frame action is,
\begin{align}
\label{bimetric2}
S_\mathrm{bi} =&M_g^2\int d^4x\sqrt{-\det g}\,R^{(g)}+M_f^2\int d^4x
\sqrt{-\det f}\,R^{(f)} \nonumber \\
&+2m^2 M_\mathrm{eff}^2 \int d^4x\sqrt{-\det g} \left( 3 - \tr \sqrt{g^{-1} f}
+ \det \sqrt{g^{-1} f} \right)\, .
\end{align}
In addition we shall assume that both the metrics $g_{\mu\nu}$ and $f_{\mu\nu}$ 
describe a flat FRW background, and by using the conformal time $t$, the 
metrics $g_{\mu\nu}$ and $f_{\mu\nu}$, are equal to,
\be
\label{Fbi10}
ds_g^2 = \sum_{\mu,\nu=0}^3 g_{\mu\nu} dx^\mu dx^\nu
= a(t)^2 \left( - dt^2 + \sum_{i=1}^3 \left( dx^i \right)^2\right) \, ,\quad
ds_f^2 = \sum_{\mu,\nu=0}^3 f_{\mu\nu} dx^\mu dx^\nu
= - c(t)^2 dt^2 + b(t)^2 \sum_{i=1}^3 \left( dx^i \right)^2 \, .
\ee
We also assume that the space-time is equipped with a symmetric, torsion-less, 
and
metric compatible affine connection, the Levi-Civita connection. By defining 
the Hubble rates for the scale factors $a(t)$, $b(t)$ and $c(t)$, to be,
\begin{equation}\label{hubblerates}
H(t)= \frac{\dot {a}(t)}{a(t)}\, ,\quad K(t)=\frac{\dot 
{b}(t)}{b(t)}\, , \quad L(t)=\frac{\dot {c}(t)}{c(t)}\, ,
\end{equation}
we obtain the following relations which constraint the final form of the Hubble 
rates,
\be
\label{identity3}
cH = bK\ \mbox{or}\
\frac{c\dot a}{a} = \dot b\, .
\ee
If $\dot a \neq 0$, we obtain $c= a\dot b / \dot a$ and on the other hand if 
$\dot a = 0$, we find $\dot b=0$, that is, $a$ and $b$ must be constant and $c$ 
can be arbitrarily chosen. Moreover, by redefining the scalar fields to be 
functions of $\eta$ and $\zeta$, that is, $\varphi=\varphi(\eta)$ and $\xi = 
\xi (\zeta)$ and also by identifying $\eta$ and $\zeta$ with the conformal time 
$t$, that is, $\eta=\zeta=t$, we obtain the following equations,
\begin{align}
\label{Fbi19}
\omega(t) M_g^2 =&  -4M_g^2 \left ( \dot{H}-H^2  \right )-2m^2 
M^2_\mathrm{eff}(ab-ac)  \, , \\
\label{Fbi20}
\tilde V (t) a(t)^2 M_g^2 =&
M_g^2 \left (2 \dot{H}+4 H^2 \right ) +m^2 M^2_\mathrm{eff}(6a^2-5ab-ac) \, , 
\\
\label{Fbi21}
\sigma(t) M_f^2 =&  - 4 M_f^2 \left ( \dot{K} - LK  \right )
   - 2m^2 M_\mathrm{eff}^2 \left ( - \frac{c}{b} + 1 \right ) \frac{a^3c}{b^2}\, , \\
\label{Fbi22}
\tilde U (t) c(t)^2 M_f^2 =&
M_f^2 \left ( 2 \dot{K} + 6 K^2 -2 L K  \right )
+ m^2 M_\mathrm{eff}^2 \left( \frac{a^3c}{b^2} - 2 c^2 + \frac{a^3c^2}{b^3} 
\right)\, ,
\end{align}
where the functions $\omega (\eta)$, $\tilde V(\eta)$, $\sigma (\eta)$ and 
$\tilde U(\eta)$, have the following dependence as functions of the variable 
$\eta$,
\be
\label{Fbi23}
\omega(\eta) = 3 \varphi'(\eta)^2 \, ,\quad
\tilde V(\eta) = V\left( \varphi\left(\eta\right) \right)\, ,\quad
\sigma(\zeta) = 3 \xi'(\zeta)^2 \, ,\quad
\tilde U(\zeta) = U \left( \xi \left(\zeta\right) \right) \, .
\ee
In effect, any arbitrarily chosen cosmological evolution of the Universe, which 
is given by specifying the scale factors $a(t)$, $b(t)$, and $c(t)$, can be 
reproduced by suitably choosing the functions $\omega(t)$, $\tilde V(t)$, 
$\sigma(t)$, and $\tilde U(t)$ in order these satisfy the equations 
(\ref{Fbi19}-\ref{Fbi22}), taking into account that Eq.~(\ref{identity3}) is 
also satisfied. In the rest of this paper we shall extensively use the 
formalism we presented in this section, in order to provide an $F(R)$ bigravity 
description of Type IV singular cosmological evolutions.

\section{General Singular Inflation from Bigravity}

\subsection{Finite-time Singularities in Cosmology}

In this section we briefly review some essential information on finite-time 
singularities, which were first classified in Ref.~\cite{Nojiri:2005sx}. The 
classification scheme used in Ref.~\cite{Nojiri:2005sx}, uses three physical 
quantities that can be consistently defined on any spacelike three dimensional 
hypersurface of constant $t$, namely the effective energy density, the 
effective
pressure, the scale factor and in addition in one case it also uses the Hubble 
rate and  its higher
derivatives. All the finite-time singularities are timelike singularities, and 
are classified as follows,
\begin{itemize}
\item Type I (``Big Rip Singularity''): It is the most phenomenologically 
``harmful''
since as the cosmic time $t$ approaches the time instance $t_s$, that is, 
$t\rightarrow t_s$, 
all the physical quantities we mentioned earlier, namely, the scale 
factor $a$, the effective energy
density $\rho_{\mathrm{eff}}$ and also the effective pressure
$p_\mathrm{eff}$ become singular at the time instance $t=t_s$, that is, $a \to 
\infty$,
$\rho_\mathrm{eff} \to \infty$, and $\left|p_\mathrm{eff}\right| \to
\infty$ respectively. We refer the reader to Refs.~\cite{ref5} for more details 
on the Big Rip singularity.
\item Type II (``Sudden Singularity''): For the Type II, as $t \to t_s$, both
the effective energy density and the scale factor are finite, that is, $a
\to a_s$, $\rho_{\mathrm{eff}}\to \rho_s$, but the effective pressure
becomes singular at $t=t_s$, $\left|p_\mathrm{eff}\right| \to \infty$. We refer 
the reader to Refs.~\cite{Barrow:2004hk,barrow}, for more 
information on this singularity.
\item Type III: In the Type III case, as $t\rightarrow t_s$, both the effective 
pressure and
the effective energy density diverge, $\rho_\mathrm{eff} \to \infty$ and
$\left|p_\mathrm{eff}\right| \to \infty$, but the scale factor does not 
diverge, that is, $a \to a_s$.
\item Type IV: This singularity is the one which concern us in this paper, 
since it is the most ``mild'' regarding the phenomenological implications. In 
this case all the aforementioned physical quantities are finite at $t=t_s$, 
that is, $a \to a_s$,
$\rho_\mathrm{eff} \to \rho_s$,
$\left|p_\mathrm{eff}\right| \to p_s$, but the higher derivatives  of order 
$n\geq 2$ of the Hubble rate diverge, but the Hubble rate is finite of course. 
We refer the reader to \cite{Barrow:2015ora,noo1,noo2,noo3,noo4} for some 
recent studies on the cosmological implications of the Type IV singularity.
\end{itemize}
As we already mentioned, the focus in this paper will be on the Type IV
singularity, which we study in the Jordan frame.

\subsection{Singular Inflation: The Bigravity Description}

Now we proceed to find how singular inflation can be realized in bigravity. 
First note that, as we already discussed before Eq.~(\ref{total}), the physical 
metric, which is determined in the frame where the scalar does not directly 
couple with matter, is given by multiplying the scalar field and the metric in 
the Einstein frame in Eq.~(\ref{Fbi10}),
\be
\label{Fbi30b}
g^\mathrm{J}_{\mu\nu} = \e^{\varphi} g_{\mu\nu}\, ,
\ee
and hereafter we shall call $g^\mathrm{J}_{\mu\nu}$ the Jordan frame metric. By 
using the cosmological time $\tilde t$ in the Jordan frame, the FRW metric is 
assumed to be equal to,
\be
\label{AA1}
ds^2 = - d\tilde t^2 + \tilde a(\tau)^2 \sum_{i=1}^3 \left( dx^i \right)^2\, .
\ee
In order to have a Type IV singular evolution, we assume that the Hubble rate 
$\tilde H (\tilde t) \equiv \frac{1}{\tilde a (\tilde t)} \frac{d \tilde 
a(\tilde t)}{d\tilde t}$, behaves as follows,
\be
\label{AA2}
\tilde H (\tilde t) \sim H_0 + H_1 \left| \tilde t - \tilde t_s 
\right|^\gamma\, .
\ee
Then, according to classification of finite-time singularities we presented 
earlier, when $\gamma\leq -1$, the cosmological evolution develops a Type I 
singularity, which is nothing but the Big Rip singularity \cite{ref5}, while 
when $-1<\gamma<0$, this case corresponds to a Type III singularity. When 
$0<\gamma<1$, the cosmological evolution develops a Type II singularity, and 
finally, in the case that $1<\gamma$ and $\gamma$ is assumed to be a 
non-integer number, this leads to a Type IV finite-time singularity. It is the 
last case that we are interested in, so hereafter we assume that $\gamma >1$. 
Notice that when $\gamma \neq -1$, the scale factor $\tilde a(\tilde t)$ 
corresponding to (\ref{AA2}) is given by,
\be
\label{AA3}
a(\tilde t) \propto \e^{ H_0 t + \sign \left( \tilde t - \tilde t_s 
\right)\frac{H_1}{\gamma + 1 }
\left| \tilde t - \tilde t_s \right|} \, .
\ee
In the above equation, the symbol $\sign$ is defined as follows,
\be
\label{AA4}
\sign (x) = \left\{ \begin{array}{cc} 1 & (x>0) \\
-1 & (x<0) \end{array}
\right.\, .
\ee
Consequently, Eq.~(\ref{AA3}) indicates that when $\gamma>-1$, the scale factor 
is finite even at the point of the singularity $\tilde t= \tilde t_s$, and this 
also covers the Type IV case. Therefore, even if we consider the conformal time 
$t$, the FRW metric is given by,
\be
\label{AA5}
ds^2 = a_J (t)^2 \left( - d t^2 + \tilde a(\tau)^2 \sum_{i=1}^3 \left( dx^i 
\right)^2\right)\, ,
\ee
and in effect, we find that $t \sim \tilde t$ owing to the fact that $d\tilde t 
= \tilde a (\tilde t) dt$ and also since $\tilde a (\tilde t) \sim 1$ at the 
singularity $\tilde t= \tilde t_s$. By looking at Eq.~(\ref{AA5}), we conclude 
that,
\be
\label{AA6}
a_J(t) = \tilde a (\tilde t)\, .
\ee
and this indicates that near the singularity, the Hubble rate in terms of 
$a_J(t)$ and of the conformal time $t$ behaves as in (\ref{AA3}), that is, \be
\label{AA7}
H_J (t) \equiv \frac{1}{a_J(t)} \frac{d a_J(t)}{dt} \sim H_0 + H_1 \left| t - 
t_s \right|^\gamma\, .
\ee
We should also note that,
\be
\label{AA8}
H_J (t) = H(t) + \frac{\dot\varphi(t)}{2}\, .
\ee
with the $H(t)$ appearing in Eq.~(\ref{AA8}), being defined after 
Eq.~(\ref{Fbi10}). Owing to the fact that the space-time is described by the 
metric $g^\mathrm{J}_{\mu\nu}$, the functions $a(t)$, $b(t)$ and $c(t)$ are not 
always directly related with the expansion of the Universe. Consequently, we 
may choose the scale factors $a(t)$, $b(t)$ and $c(t)$ in a way consistent with 
Eq.~(\ref{identity3}). In the following, we shall assume that $a(t)=b(t)=1$, so 
that, Eq.~(\ref{AA8}) indicates that,
\be
\label{AA9}
H_J (t) = \frac{\dot\varphi(t)}{2}\, .
\ee
Then Eqs.~(\ref{Fbi19}), (\ref{Fbi20}), (\ref{Fbi21}), and (\ref{Fbi22}) are 
simplified as follows,
\begin{align}
\label{Fbi19C}
\omega (t)^2 M_g^2 =& 12 M_g^2 H_J^2 = m^2 M_\mathrm{eff}^2
\left( c - 1\right) \, , \\
\label{Fbi20C}
\tilde V (t) M_g^2 =& m^2 M_\mathrm{eff}^2 \left( 1 - c \right)
= - 6 M_g^2 H_J^2 \, , \\
\label{Fbi21C}
\sigma(t) M_f^2 =& 2 m^2 M_\mathrm{eff}^2 \left( c - 1 \right)
= 12 M_g^2 H_J^2 \, , \\
\label{Fbi22C}
\tilde U (t) M_f^2 =& m^2 M_\mathrm{eff}^2 c \left( 1 - c \right)
= - 6 M_g^2 H_J^2
\left( 1 + \frac{ 6 H_J^2}{m^2 M_\mathrm{eff}^2} \right) \, .
\end{align}
The Eq.~(\ref{Fbi19C}) can be solved explicitly with respect to $c$, and it 
yields,
\be
\label{FFFbi1}
c (t)= 1 + \frac{ 6 H_J^2}{m^2 M_\mathrm{eff}^2} \, .
\ee
Then, by using (\ref{AA7}), we find that the model generating Type II, III, or 
IV singularities is given by
\begin{align}
\label{Fbi19D}
\omega (t)^2 M_g^2 =& 12 M_g^2 \left( H_0 + H_1 \left| t - t_s \right|^\gamma 
\right)^2 \, , \\
\label{Fbi20D}
\tilde V (t) M_g^2 =& - 6 M_g^2 \left( H_0 + H_1 \left| t - t_s \right|^\gamma 
\right)^2\, , \\
\label{Fbi21D}
\sigma(t) M_f^2 =& 12 M_g^2 \left( H_0 + H_1 \left| t - t_s \right|^\gamma 
\right)^2 \, , \\
\label{Fbi22D}
\tilde U (t) M_f^2 =& - 6 M_g^2 \left( H_0 + H_1 \left| t - t_s \right|^\gamma 
\right)^2
\left( 1 + \frac{ 6 \left( H_0 + H_1 \left| t - t_s \right|^\gamma \right)^2
}{m^2 M_\mathrm{eff}^2} \right) \, .
\end{align}
and therefore we find that effectively, $c(t)$ is equal to,
\be
\label{FFFbi1B}
c (t) = 1 + \frac{ 6 \left( H_0 + H_1 \left| t - t_s \right|^\gamma 
\right)^2}{m^2 M_\mathrm{eff}^2} \, .
\ee
Let us now proceed to the calculation of the slow-roll parameters for the Type 
IV singular evolution model, and we shall calculate in detail the slow-roll 
parameters $\epsilon$, $\eta$ and $\xi$. When we use the cosmological time $t$ 
in (\ref{AA1}) and the e-foldings $N$ defined by $a=a_0\e^N$ with a constant 
$a_0$, we can express the slow-roll parameters by using $\tilde H$ as follows,
\begin{align}
\label{S7}
\epsilon
=& - \frac{\tilde H(N)}{4 \tilde H'(N)} \left[ \frac{6\frac{\tilde 
H'(N)}{\tilde H(N)}
+ \frac{\tilde H''(N)}{\tilde H(\phi)} + \left( \frac{\tilde H'(N)}{\tilde 
H(N)} \right)^2}
{3 + \frac{\tilde H'(N)}{\tilde H(N)}} \right]^2 \, , \nn
\eta = & -\frac{1}{2} \left( 3 + \frac{\tilde H'(N)}{\tilde H(N)} \right)^{-1} 
\left[
9 \frac{\tilde H'(N)}{\tilde H(N)} + 3 \frac{\tilde H''(N)}{\tilde H(N)}
+ \frac{1}{2} \left( \frac{\tilde H'(N)}{\tilde H(N)} \right)^2 -\frac{1}{2} 
\left( \frac{\tilde H''(N)}{\tilde H'(N)} \right)^2
+ 3 \frac{\tilde H''(N)}{\tilde H'(N)} + \frac{\tilde H'''(N)}{\tilde H'(N)} 
\right] \, , \nn
\xi^2 = & \frac{ 6 \frac{\tilde H'(N)}{\tilde H(N)} + \frac{\tilde 
H''(N)}{\tilde H(N)}
+ \left( \frac{\tilde H'(N)}{\tilde H(N)} \right)^2 }{4 \left( 3 + \frac{\tilde 
H'(N)}{\tilde H(N)} \right)^2}
\left[ 3 \frac{\tilde H(N) \tilde H'''(N)}{\tilde H'(N)^2} + 9 \frac{\tilde 
H'(N)}{\tilde H(N)}
  - 2 \frac{\tilde H(N) \tilde H''(N) \tilde H'''(N)}{\tilde H'(N)^3} + 4 
\frac{\tilde H''(N)}{\tilde H(N)}
\right. \nn
& \left.
+ \frac{\tilde H(N) \tilde H''(N)^3}{\tilde H'(N)^4} + 5 \frac{\tilde 
H'''(N)}{\tilde H'(N)}
  - 3 \frac{\tilde H(N) \tilde H''(N)^2}{\tilde H'(N)^3} - \left( \frac{\tilde 
H''(N)}{\tilde H'(N)} \right)^2
+ 15 \frac{\tilde H''(N)}{\tilde H'(N)}
+ \frac{\tilde H(N) \tilde H''''(N)}{\tilde H'(N)^2} \right]\, .
\end{align}
In addition, the relations between $\tilde H(N)$ and $H_J(t)$ are given below,
\begin{align}
\label{AA10}
\tilde H(N) =& \frac{H_J(t)}{a(t)}\, , \quad \tilde H'(N) = \frac{1}{a(t) 
H_J(t)} \left( - H_J(t)^2 + \dot H_J(t) \right)\, , \nn
\tilde H(N)'' =& \frac{1}{a(t) H_J(t)} \left( H_J(t)^2 - 2 \dot H_J(t) - 
\frac{\left( \dot H_J(t) \right)^2}{H_J(t)^2} + \frac{ \ddot H_J(t)}{H_J(t)^2} 
\right)\, , \nn
\tilde H(N)''' =& \frac{1}{a(t) H_J(t)} \left( - H_J(t)^2 +5 \dot H_J(t) - 
\frac{\left( \dot H_J(t) \right)^2}{H_J(t)^2} - \frac{ 3 \ddot 
H_J(t)}{H_J(t)^2}
  - \frac{2 \dot H_J(t) \ddot H_J(t)}{H_J(t)^3} + \frac{\dddot H_J(t)}{H_J(t)^2} 
\right)\, , \nn
\tilde H(N)'''' =& \frac{1}{a(t) H_J(t)} \left( H_J(t)^2 -8 \dot H_J(t) + 
\frac{6 \left( \dot H_J(t) \right)^2}{H_J(t)^2}
+ \frac{\left( \dot H_J(t) \right)^3}{H_J(t)^3} + \frac{ 8 \ddot 
H_J(t)}{H_J(t)^2} - \frac{ 2 \left( \ddot H_J(t) \right)^2}{H_J(t)^4} \right. 
\nn
& \left. + \frac{2 \left( \dot H_J(t) \right)^2 \ddot H_J(t)}{H_J(t)^5} - 
\frac{4 \dddot H_J(t)}{H_J(t)^2}
  - \frac{3 \dot H_J(t) \dddot H_J(t)}{H_J(t)^4}  + \frac{\ddddot 
H_J(t)}{H_J(t)^2} \right)\, .
\end{align}
Having these at hand, we may evaluate the slow-roll parameters near the Type IV 
singularity at $t \sim t_s$. Recall that since the singularity is a Type IV 
one, the parameter $\gamma $ has to obey $\gamma>1$ and also $\gamma$ must not 
an integer in (\ref{AA7}). Therefore, by using the above equations, when 
$1<\gamma<2$, we find,
\begin{align}
\label{AA11}
& \tilde H \sim \frac{H_0}{a\left(t_s \right)}\, , \quad \tilde H' \sim - 
\frac{H_0}{a\left(t_s \right)}\, , \quad
\tilde H'' \sim \frac{H_1 \gamma \left( \gamma -1 \right)}{a\left(t_s \right) 
H_0^3} \left| t - t_s \right|^{\gamma-2} \, , \nn
& \tilde H''' \sim \frac{H_1 \gamma \left( \gamma -1 \right) \left( \gamma -2 
\right)} \sign \left( t - t_s \right)
{a\left(t_s \right) H_0^3} \left| t - t_s \right|^{\gamma-3} \, , \quad
\tilde H'''' \sim \frac{H_1 \gamma \left( \gamma -1 \right) \left( \gamma -2 
\right) \left( \gamma -3 \right)}
{a\left(t_s \right) H_0^3} \left| t - t_s \right|^{\gamma-4} \, .
\end{align}
In the case that $2<\gamma<3$, the term $\tilde H''$ in (\ref{AA1}) is replaced 
by the following expression,
\be
\label{AA12}
\tilde H'' \to \frac{H_0}{a\left(t_s \right)}\, ,
\ee
while in the case that $3<\gamma<4$, in addition to the term $\tilde H''$, also 
$\tilde H'''$ should be taken into account, and it is replaced by,
\be
\label{AA13}
\tilde H''' \to - \frac{H_0}{a\left(t_s \right)}\, .
\ee
Finally, when $\gamma > 4$ and $\gamma$ is not an integer, the term $\tilde 
H''''$ is replaced by the following expression,
\be
\label{AA14}
\tilde H'''' \to \frac{H_0}{a\left(t_s \right)}\, .
\ee
Consequently, when $1<\gamma<2$, we find that the slow-roll parameters become,
\begin{align}
\label{AA15}
& \epsilon \sim - \frac{H_1 \gamma \left( \gamma - 1\right)}{16 H_0^4} \left| t 
- t_s \right|^{\gamma -2}\, , \quad
\eta \sim \frac{H_1 \gamma \left( \gamma - 1\right) \left( \gamma - 2\right)}{4 
H_0^4} \sign \left( t - t_s \right) \left| t - t_s \right|^{\gamma -3}\, , \nn
&\xi^2\sim - \frac{5 H_1 \gamma \left( \gamma - 1\right) \left( \gamma - 
2\right) \left( \gamma - 3\right)}{16 H_0^4} \left| t - t_s \right|^{\gamma 
-4}\, ,
\end{align}
which are singular at $t=t_s$. In the case $2<\gamma<3$, $\epsilon$ becomes 
finite even at $t=t_s$ and we find that the slow-roll parameter is equal to,
\be
\label{AA16}
\epsilon = - \frac{1}{4}\, .
\ee
In the case $3<\gamma<4$, the slow-roll parameter $\eta$ also becomes finite 
and it is equal to,
\be
\label{AA17}
\eta = 2 \, .
\ee
In addition, when $\gamma>4$ and $\gamma$ is not an integer, all of $\epsilon$, 
$\eta$, and $\xi^2$ are finite and we find that,
\be
\label{AA18}
\xi^2 = 4\, .
\ee
Hence, we demonstrated that the slow-roll indices may be singular in the case 
that a Type IV singularity occurs in the cosmological evolution. But what does 
exactly this singularity indicates? As was shown in Refs.~\cite{noo4}, 
singularities in the slow-roll indices indicate dynamical instability of the 
inflationary process. This does not by no mean that the spectral observational 
indices of inflation become infinite. As was demonstrated in 
\cite{barrowslowroll}, the slow-roll indices $\epsilon $ and $\eta$ are first 
order terms in the slow-roll expansion. So when these become large (of order 
$\sim 1$), the slow-roll expansion breaks down and inflation ends. This is 
exactly what happens in our case, and since the slow-roll parameters diverge at 
$t=t_s$, this means that inflation should end there in an abrupt way. We shall 
further discuss this issue in a later section, in more detail, but for 
illustrative reasons, let us assume that we calculate the inflationary 
observational indices, and we assume that we use the singular expressions for 
the slow-roll parameters we found in this section. For simplicity, let us 
assume that the definition of the inflationary indices is that of a canonical 
scalar coupled with gravity, so the spectral index of the primordial curvature 
fluctuations $n_s$, the tensor-to-scalar ratio $r$, and the associated running 
of the spectral index $\alpha_s$ are given by
\be
\label{AAA1}
n_s \sim 1 - 6 \epsilon + 2 \eta\, , \quad r \sim 16\epsilon\, , \quad
\alpha_s \sim 16\epsilon \eta - 24 \epsilon^2 - 2 \xi^2\, .
\ee
Although there is no reason that the expressions in (\ref{AAA1}) can be 
justified for the $F(R)$ bigravity model in this paper, we may evaluate the 
quantities $n_s$, $r$, and $\alpha_s$ by using (\ref{AAA1}), by assuming the 
end of the inflation is given by $t=t_f$ but also that $t_f \sim t_s$. Then, in 
the case that $1<\gamma<2$, we find,
\begin{align}
\label{AAA2}
& n_s \sim \frac{H_1 \gamma \left( \gamma - 1\right) \left( \gamma - 
2\right)}{2 H_0^4} \sign \left( t_f - t_s \right) \left| t_f - t_s 
\right|^{\gamma -3}\, , \quad
r \sim  - \frac{H_1 \gamma \left( \gamma - 1\right)}{H_0^4} \left| t_f - t_s 
\right|^{\gamma -2}\, , \nn
& \alpha_s \sim \frac{5 H_1 \gamma \left( \gamma - 1\right) \left( \gamma - 
2\right) \left( \gamma - 3\right)}{8 H_0^4} \left| t_f - t_s \right|^{\gamma 
-4}\, ,
\end{align}
and correspondingly in the case that $2<\gamma<3$, we obtain,
\begin{align}
\label{AAA3}
& n_s \sim \frac{H_1 \gamma \left( \gamma - 1\right) \left( \gamma - 
2\right)}{2 H_0^4} \sign \left( t_f - t_s \right) \left| t_f - t_s 
\right|^{\gamma -3}\, , \quad
r \sim  - 4 \, , \nn
& \alpha_s \sim \frac{5 H_1 \gamma \left( \gamma - 1\right) \left( \gamma - 
2\right) \left( \gamma - 3\right)}{8 H_0^4} \left| t_f - t_s \right|^{\gamma 
-4}\, .
\end{align}
Furthermore, in the case of $3<\gamma<4$, we obtain,
\be
\label{AAA4}
n_s \sim \frac{13}{2} \, , \quad
r \sim  - 4 \, , \quad
\alpha_s \sim \frac{5 H_1 \gamma \left( \gamma - 1\right) \left( \gamma - 
2\right) \left( \gamma - 3\right)}{8 H_0^4} \left| t_f - t_s \right|^{\gamma 
-4}\, .
\ee
and finally, in the case that $\gamma>4$, we get,
\be
\label{AAA5}
n_s \sim \frac{13}{2} \, , \quad
r \sim  - 4 \, , \quad
\alpha_s \sim - \frac{35}{2} \, .
\ee
By looking at Eqs.~(\ref{AAA2}), (\ref{AAA3}), and (\ref{AAA4}), we may loosely 
say that the spectral inflationary indices may diverge at the singularity, but 
this is not true. This is owing to the fact that the spectral indices are given 
in terms of the slow-roll parameters only in the case the slow-roll 
approximation holds true, so when $\epsilon,\eta \ll 1$, which is not the case 
when a singularity appears in the slow-roll parameters. It can be true that 
$\epsilon,\eta \ll 1$ long before the singularity, but at the singularity, the 
inflationary indices cannot be written in terms of the slow-roll parameters 
since the perturbation slow-roll expansion breaks down at the singularity.

Notice, however, that in the case of $1<\gamma<2$, by suitably choosing the 
parameters $H_0$, $H_1$, $\gamma$, and $t_f - t_s$, the resulting expression 
can be compatible with the 2015 Planck report \cite{planck} which restricts the 
inflationary indices as follows,
\be
\label{planckconstr}
n_s=0.9644\pm 0.0049\, , \quad r<0.10\, , \quad a_s=-0.0057\pm 0.0071\, .
\ee


\section{Singular Nearly $R^2$ Inflation from Bigravity}

In the previous section we discussed how a Type IV singular cosmological 
evolution can be produced by bigravity, and as we demonstrated this leads to 
slow-roll parameters that may contain singularities. These singularities can be 
valuable from a physical point of view, since these indicate strong 
instabilities of the dynamical evolution of the physical system. In this 
section we shall study a Type IV singular version of the Starobinsky $R^2$ 
inflation model \cite{starobinsky}, to which we refer as ``singular $R^2$ 
inflation'' model, hereafter. Particularly, we shall incorporate the Type IV 
singularity in the Hubble rate that corresponds to the Jordan frame $R^2$ 
inflation model, the action of which is,
\begin{equation}
\label{action1dse}
\mathcal{S}=\frac{1}{2\kappa^2}\int \mathrm{d}^4x\sqrt{-g}\left( 
R+\frac{1}{6M^2}R^2\right)\, ,
\end{equation}
with the parameter satisfying $M\gg 1$. Note that our approach is not directly 
related to the Jordan frame $F(R)$ theory, since we shall try to generate the 
Hubble rate that the vacuum $F(R)$ theory of Eq.~(\ref{action1dse}) generates 
by using the bigravity theoretical framework. To this end, let us calculate the 
Hubble rate that is generated by the $F(R)$ gravity of Eq.~(\ref{action1dse}), 
so that we use it as a basis for our calculations. The FRW equation 
corresponding to the $F(R)$ gravity of Eq.~(\ref{action1dse}) is equal to,
\begin{equation}\label{takestwo}
\ddot{H}-\frac{\dot{H}^2}{2H}+\frac{M^2}{2}H=-3H\dot{H}\, ,
\end{equation}
and owing to the fact that during inflation, the terms $\ddot{H}$ and $\dot{H}$ 
can be considered subdominant, the resulting Hubble rate can be found by 
solving the differential equation of Eq.~(\ref{takestwo}), and the result is 
approximately equal to,
\begin{equation}\label{hubstar}
H(t)\simeq H_i-\frac{M^2}{6}\left ( t-t_i\right )\, .
\end{equation}
In the above equation, $t_i$ is the initial time instance that we assume 
inflation starts and the parameter $H_i$ is the corresponding value of the 
Hubble rate at the time instance $t=t_i$.

The purpose of this section is twofold, firstly we shall see how a singular 
$R^2$ inflation can be generated by a bigravity theory and secondly we shall 
see what are the new qualitative features of this singular $R^2$ inflation 
model, focusing on the inflationary indices. As we shall demonstrate, the 
``harmless'' Type IV singularity may generate strong dynamical instabilities 
manifested in the slow-roll parameters. The simplest way to add a Type IV 
singularity in the $R^2$ inflation Hubble rate of Eq.~(\ref{hubstar}), is the 
following,
\begin{equation}\label{singstarobhub}
H(t)\simeq H_i-\frac{M^2}{6}\left ( t-t_i\right )+f_0\left (t-t_s
\right)^{\gamma} \, ,
\end{equation}
where we assumed that the Type IV singularity occurs at $t=t_s$, and also that 
$\gamma>1$, in order for the cosmological evolution to develop a Type IV 
singularity. In addition, in order the effects of the singularity on the 
cosmological evolution are small, we further assume that $H_i\gg f_0$, 
$M\gg f_0$ and also that $f_0\ll 1$.
In effect, the singularity term $\sim f_0\left (t-t_s
\right)^{\gamma}$, is significantly smaller compared to the other two terms of 
Eq.~(\ref{singstarobhub}). Consequently, the effect of the singularity on the 
Hubble rate is practically insignificant, when $t\rightarrow t_s$.

Before going into the detailed calculation of the slow-roll indices, let us see 
which bigravity model can produce the cosmological evolution of 
Eq.~(\ref{singstarobhub}).
We can easily see that if we determine the functions $\omega (t)$, $V(t)$, 
$\sigma (t)$ and $U(t)$ which appear in Eq.~(\ref{Fbi19D}), which by 
substituting the Hubble rate of Eq.~(\ref{singstarobhub}) in 
Eq.~(\ref{Fbi19D}), we easily get,
\begin{align}
\label{bigravitymodel}
\omega (t)=& 2 \sqrt{3} \left(H_i-\frac{1}{6} M^2 (t-t_i)+f_0 (t-t_s)^{\gamma 
}\right) \, , \nn 
\tilde V(t)=& -6 \left(H_i-\frac{1}{6} M^2 (t-t_i)+f_0 (t-t_s)^{\gamma 
}\right)^2 \, , \nn
\sigma (t)= &\frac{12 M_g^2 \left(H_i-\frac{1}{6} M^2 (t-t_i)+f_0 (t-t_s)^{\gamma 
}\right)^2}{M_f^2} \, , \nn
\tilde U(t)=& -\frac{6 M_g^2 \left(1+\frac{6 \left(H_i-\frac{1}{6} M^2 
(t-t_i)+f_0 (t-t_s)^{\gamma }\right)^2}{m^2 M_\mathrm{eff}^2}\right) 
\left(H_i-\frac{1}{6} M^2 (t-t_i)+f_0 (t-t_s)^{\gamma }\right)^2}{M_f^2}\, .
\end{align}
and note that we took into account Eq.~(\ref{AA6}), so the above relations hold 
true near the singularity, where the conformal time is nearly identical with 
the cosmic time. We proceed to the calculation of the slow-roll parameters, and 
by using Eqs.~(\ref{S7}) and (\ref{singstarobhub}), we can obtain the exact 
form of the slow-roll indices $\epsilon$ and $\eta$, which appear in the 
Appendix. The parameter $\xi$ is very large to be presented in detail, so we 
presented in the Appendix only the terms that can be singular at the Type IV 
singularity time instance $t=t_s$. As it can be seen in the Appendix, the 
slow-roll parameters can be singular at $t=t_s$ for various values of the 
parameter $\gamma$, when $\gamma>1$. To see this explicitly, we present the 
approximate form of the slow-roll parameters for $t\rightarrow t_s$, starting 
with the parameter $\epsilon$, which reads,
\begin{equation}
\label{slowrolleps}
\epsilon\simeq -\frac{3 f_0^2 (t-t_s)^{-4+2 \gamma } (-1+\gamma )^2 \gamma 
^2}{2 \left(-6 H_i+M^2 (-1+t-t_i)\right) \left(H_i+\frac{1}{6} M^2 
(-t+t_i)\right)^5 \left(3-\frac{36 \left(6 H_i+M^2 (1-t+t_i)\right)}{\left(6 
H_i+M^2 (-t+t_i)\right)^3}\right)^2}\, .
\end{equation}
As it can be seen, when $1<\gamma<2$, the term $\sim(t-t_s)^{2(-2+\gamma)}$, 
becomes singular at $t=t_s$ and therefore the slow-roll parameter $\epsilon$ 
becomes singular too. This infinite instability clearly indicates that the 
dynamics of inflation are strongly disturbed at the infinite instability time 
instance. We shall thoroughly discuss this issue in the next section, so we 
refer from going into further details on this issue for the moment. In the 
case, $\gamma >2$, the slow-roll parameter $\epsilon$ becomes,
\begin{align}
\label{slowrollnonsingular}
\epsilon \simeq& -\frac{\left(6 H_i+M^2 (-t+t_i)\right)^2 \left(72 
M^2+\left(36+36 H_i^2-12 H_i \left(18+M^2 (t-t_i)\right)\right.\right.}{1119744 
\left(-6 H_i+M^2 (-1+t-t_i)\right) \left(H_i+\frac{1}{6} M^2 (-t+t_i)\right)^5 
\left(3-\frac{36 \left(6 H_i+M^2 (1-t+t_i)\right)}{\left(6 H_i+M^2 
(-t+t_i)\right)^3}\right)^2} \nn
& -\frac{\left.\left.12 M^2 (-2+3 t-3 t_i)+M^4 (t-t_i)^2\right) \left(6 H_i+M^2 
(-t+t_i)\right)\right)^2}{1119744 \left(-6 H_i+M^2 (-1+t-t_i)\right) 
\left(H_i+\frac{1}{6} M^2 (-t+t_i)\right)^5 \left(3-\frac{36 \left(6 H_i+M^2 
(1-t+t_i)\right)}{\left(6 H_i+M^2 (-t+t_i)\right)^3}\right)^2}\, ,
\end{align}
so the Type IV singularity has no effect on the slow-roll parameter $\epsilon$ 
in this case. Accordingly, the slow-roll index $\eta$ for the Hubble rate 
(\ref{singstarobhub}) is given in detail in the Appendix, and below we quote 
the approximate form of $\eta$ near the singularity at $t\simeq t_s$, by 
keeping only the terms that contain singularities,
\begin{align}
\label{singularityetaterms}
\eta_{sing}\simeq & \frac{1}{4 \left(3-\frac{36 \left(6 H_i+M^2 
(1-t+t_i)\right)}{\left(6 H_i+M^2 (-t+t_i)\right)^3}\right)} 
\left( -\frac{3 f_0 (t-t_s)^{-2+\gamma } (-1+\gamma ) \gamma 
}{\left(H_i+\frac{1}{6} M^2 (-t+t_i)\right)^4} \right. \nn
& -\frac{36 f_0 (t-t_s)^{-2+\gamma 
} (-1+\gamma ) \gamma }{\left(-6 H_i+M^2 (-1+t-t_i)\right) 
\left(H_i+\frac{1}{6} M^2 (-t+t_i)\right)^2}+\frac{f_0^2 (t-t_s)^{-4+2 \gamma } 
(-1+\gamma )^2 \gamma ^2}{\left(H_i+\frac{1}{6} M^2 (-t+t_i)\right)^8} \nn 
& +12 \left(\frac{72 f_0 M^2 (t-t_s)^{-2+\gamma } (-1+\gamma ) \gamma }{\left(6 
H_i+M^2 (-t+t_i)\right)^3-6 H_i+M^2 (-1+t-t_i)} \right. \nn
& \left. \left. -\frac{36 f_0 (t-t_s)^{-3+\gamma 
} (2+3 t-3 t_s-\gamma ) (-1+\gamma ) \gamma }{\left(6 H_i+M^2 
(-t+t_i)\right)^2-6 H_i+M^2 (-1+t-t_i)}\right) \right) \, .
\end{align}
In this case, the terms that contain singularities are the ones listed below,
\begin{itemize}
     \item When $1<\gamma<2$, the singular terms are, $\sim (t-t_s)^{-2+\gamma 
}$, $\sim (t-t_s)^{-3+\gamma }$, $\sim (t-t_s)^{-4+2 \gamma }$
     \item When $2<\gamma<3$, the singular terms are, $\sim (t-t_s)^{-3+\gamma 
}$
\end{itemize}
When $\gamma>3 $ no singular terms occur, so the slow-roll parameter $\eta$ 
reads,
\begin{align}
\label{smjhf}
\eta \simeq & -\frac{1}{4 \left(3-\frac{36 \left(6 H_i+M^2 
(1-t+t_i)\right)}{\left(6 H_i+M^2 (-t+t_i)\right)^3}\right)} \left( 
-\frac{648 \left(6 H_i+M^2 (1-t+t_i)\right)}{\left(6 H_i+M^2 
(-t+t_i)\right)^3}+\frac{36 \left(6 H_i+M^2 (1-t+t_i)\right)^2}{\left(6 H_i+M^2 
(-t+t_i)\right)^4} \right. \nn 
& +\frac{12 \left(-\frac{5 M^2}{6}-\left(H_i+\frac{1}{6} M^2 
(-t+t_i)\right)^2-\frac{M^4}{\left(6 H_i+M^2 (-t+t_i)\right)^2}\right)}{-6 
H_i+M^2 (-1+t-t_i)} \nn
& -\frac{\left(\frac{M^2}{3}+\left(H_i+\frac{1}{6} M^2 
(-t+t_i)\right)^2-\frac{M^4}{\left(6 H_i+M^2 
(-t+t_i)\right)^2}\right)^2}{\left(H_i+\frac{1}{6} M^2 (-t+t_i)\right)^4} \nn
& +\frac{3 \left(\frac{M^2}{3}+\left(H_i+\frac{1}{6} M^2 
(-t+t_i)\right)^2-\frac{M^4}{\left(6 H_i+M^2 
(-t+t_i)\right)^2}\right)}{\left(H_i-\frac{1}{6} M^2 (t-t_i)+f_0 
(t-t_s)^{\gamma }\right)^2} \nn
& \left. +\frac{36 \left(\frac{M^2}{3}+\left(H_i+\frac{1}{6} 
M^2 (-t+t_i)\right)^2-\frac{M^4}{\left(6 H_i+M^2 (-t+t_i)\right)^2}\right)}{-6 
H_i+M^2 (-1+t-t_i)-6 f_0 (t-t_s)^{-1+\gamma } (t-t_s-\gamma )} \right)\, .
\end{align}
As in the case of the slow-roll parameter, the singularities on the slow-roll 
index clearly indicate that the dynamics of inflation as these are quantified 
in terms of the slow-roll index $\eta$, are in some way interrupted, or more 
correctly become unstable at the singularity point. Let us recall what the 
slow-roll parameters $\epsilon$ and $\eta$ indicate for the inflationary 
dynamics. First, these parameters are of first order in the slow-roll 
perturbative expansion, with the slow-roll parameter $\epsilon$ indicating if 
inflation occurs, and the second slow-roll parameter measuring how much does 
inflation lasts \cite{barrowslowroll,noo4}. Therefore, the infinite instability 
of these parameters at some time instance clearly indicates that the dynamics 
is interrupted. However, the dynamics might also be interrupted at higher 
orders in the Hubble slow-roll expansion, therefore this could be an indication 
that inflation ends at the singularity. This issue is very important and needs 
to be further discussed, so we defer this discussion to the next section. 
Before closing this section, let us note that we calculated the slow-roll 
parameter $\xi$, for the Hubble evolution of Eq.~(\ref{singstarobhub}), but the 
resulting expression is too large to quote it here, and even the approximate expression 
near the singularity is quite lengthy. So in the Appendix we have included all 
the terms that contain singularities in the case of the slow-roll parameter 
$\xi$. Below we quote the singular terms that the parameter $\xi$ contains, for 
various values of the parameter $\gamma$.
\begin{itemize}
     \item When $1<\gamma<2$, the singular terms are, $\sim (t-t_s)^{-6+3 \gamma 
}$, $(t-t_s)^{-4+\gamma }$, $\sim (t-t_s)^{-2+\gamma }$, $\sim 
(t-t_s)^{-3+\gamma }$, $\sim (t-t_s)^{-4+2 \gamma }$
     \item When $2<\gamma<3$, the singular terms are, $\sim (t-t_s)^{-3+\gamma 
}$, $(t-t_s)^{-4+\gamma }$
     \item When $3<\gamma <4$, the only singular term is $(t-t_s)^{-4+\gamma }$.
\end{itemize}
It is conceivable that the parameter $\xi$ is free of singularities when 
$\gamma >4$. In Table~\ref{newtab}, we gathered all the results, regarding the 
singularities of the slow-roll parameters at $t=t_s$, for various values of the 
parameter $\gamma$. As it can be seen, when $\gamma>4$, the slow-roll 
parameters are free of singularities. As a concluding remark, we need to note 
that although that the singularity remains unnoticed when one considers the 
Hubble rate and all the physical quantities that can be defined on the 
three-dimensional spacelike hypersurface $t=t_s$, it has a strong effect on the 
dynamics of the cosmological evolution, drastically affecting the slow-roll 
parameters, which control the dynamics of inflation.

\begin{table*}[h]
\small
\caption{\label{newtab}Singularity Structure of the slow-roll parameters 
$\epsilon$, $\eta$ and $\xi$ for various values of the parameter $\gamma$}
\begin{tabular}{@{}ccccccc@{}}
\tableline
\tableline
\tableline
Slow-roll Parameter & \ \qquad \ & $1<\gamma<2$ & \ \qquad \ & $2<\gamma<3$ & \ \qquad \ & $3<\gamma<4$ \\
\tableline
$\epsilon$ && Singular && Non-singular && Non-singular \\
\tableline
$\eta$ && Singular &&Singular && Non-singular \\
\tableline
$\xi$ && Singular && Singular && Singular \\
\tableline
\tableline
\tableline
\end{tabular}
\end{table*}


\subsection{Graceful Exit via Dynamical Instabilities-A Critical Discussion}

In the previous two sections we demonstrated how a Type IV singularity can be 
incorporated in the cosmological evolution of the Universe, and we investigated 
which bigravity theory can successfully generate such an evolution. 
Particularly we studied two cosmological models, one corresponding to the 
Hubble rate of Eq.~(\ref{AA2}) and another one with the Hubble rate being the 
one of Eq.~(\ref{singstarobhub}). The latter model is a singular deformation of 
the $R^2$ inflation model \cite{starobinsky}. In both cases, the Type IV 
singularity in the Hubble rate is generated by a term $\sim (t-t_s)^{\gamma}$, 
where the parameter $\gamma$ is assumed to be $\gamma>1$, so that a Type IV 
singularity occurs.  As we showed, the presence of the singularity plays no 
role in the cosmological evolution, since the Hubble rate and all the physical 
quantities that can be defined on the three dimensional spacelike hypersurface 
$t=t_s$ are finite. However, the effects of the singularity are quite severe 
when the slow-roll parameters are taken into account. Indeed, by calculating 
these, we demonstrated that these become infinite for certain values of the 
parameter $\gamma$. Recall that the slow-roll parameters determine the dynamics 
of inflation \cite{barrowslowroll}, and actually these indicate if inflation 
occurs in the first place and how long it lasts. Particularly, the slow-roll 
parameter $\epsilon$ determines if inflation begins, and in order slow-roll 
evolution occurs, it must obey $\epsilon\ll 1$, while the parameter $\eta $, 
determines how long inflation lasts. The presence of singularities in these 
slow-roll parameter clearly indicates that the dynamics of inflation becomes 
unstable at the singularity point. This infinite instability shows that the 
dynamical evolution is abruptly interrupted at the singularity and therefore 
the inflationary attractor that described the dynamical system up to that point 
ceases to be the final attractor of the theory. Consequently, the presence of 
singularities indicates that at the point of singularities, the graceful exit 
from inflation occurs. Of course it is conceivable that the singularities 
solely do not generate graceful exit from inflation, but they provide clear 
information that the graceful exit occurs at that point. The actual mechanism 
for graceful exit should be some curvature perturbation instability 
\cite{sergeitraceanomaly,noo4} or a tachyonic instability \cite{tachyon}, but 
nevertheless, this graceful exit should occur at the singularity point $t=t_s$. 
For some relevant studies of the infinite instability that the Type IV 
singularity generates, see Ref.~\cite{noo4}. We need to stress that the 
slow-roll parameters $\epsilon$ and $\eta$, are of first order in the slow-roll 
perturbative expansion \cite{barrowslowroll}, and therefore the presence of 
instabilities at these parameters generates the question whether inflation 
actually ended at a higher order in the perturbative slow-roll expansion. 
Actually, if higher order slow-roll parameters also become infinite at the Type 
IV singularity, this shows that the perturbative slow-roll expansion breaks at 
higher order and hence, this indicates that inflation ends since the 
perturbative expansion breaks at a higher order. In principle we could go 
towards this direction and calculate these higher order slow-roll parameters, 
but their complicated form would make the presentation of the paper unnecessary 
complicated, so we defer from going into detail. But it can be easily checked 
that the qualitative picture we just discussed, indeed holds true.

Before closing this section, we need to stress another interesting possibility, 
related to the singular $R^2$ Starobinsky inflation model. In the ordinary 
Starobinsky model, graceful exit from inflation occurs when the slow-roll 
parameter $\epsilon$ becomes of the order $\sim 1$, and suppose that this 
happens when $t=t_f$. As it can be checked, the second slow-roll parameter 
$\eta$ is finite at the moment that $\epsilon$ becomes of the order $\sim 1$ 
\cite{noo4}, that is, at the time instance $t=t_f$.  In the singular $R^2$ 
model, regardless when $\epsilon$ becomes of the order $\epsilon \sim 1$, at the singularity, the slow-roll 
parameters become infinite and therefore inflation might end at the singularity 
in a more abrupt way, since the inflationary dynamics are severely interrupted 
at the singular point. Hence, we may have two interesting possibilities, either 
inflation ends at $t=t_s$, with $t_s<t_f$, and therefore earlier from the time 
instance $t=t_f$, or inflation ends at $t=t_f=t_s$, so in this case inflation 
ends at the time that the ordinary $R^2$ inflation exits from inflation, with 
the difference being that, in the singular Starobinsky case, inflation ends more 
abruptly. For a thorough discussion on these issues, we refer the reader to 
Refs.~\cite{noo4}. Finally, let us note that in principle someone would claim 
that the observational indices $n_s$ and $r$ become infinite at the 
singularities, this however is not true. This is owing to the fact that the 
spectral index of primordial curvature perturbations can be written in terms of 
the slow-roll parameters $\epsilon$ and $\eta$ only in the case that these 
satisfy the slow-roll condition $\epsilon, \eta \ll 1$. In addition, these are 
calculated at the time that the quantum fluctuations of the comoving scalar 
curvature exit the horizon, that is, when the corresponding wavelengths become 
of the order of the Hubble radius $r_h=\frac{1}{a(t)H(t)}$, which occurs much 
more earlier than the graceful exit from inflation. Therefore, no infinity can 
occur at the observational indices and the interpretation of the singularities 
appearance in the slow-roll indices clearly affects only the graceful exit from 
inflation era. So the corresponding wavelengths that exit the horizon at the 
moment that graceful exit occurs, are irrelevant to present time observations. 
This is because the only modes that are relevant at present time are the ones 
with wavelength equal to the Hubble radius at the moment of horizon crossing 
long before the graceful exit, which re-enter the horizon after the reheating 
of the Universe. Hence, no infinities at the observational indices occur, and these occur only 
at the slow-roll parameters As we evinced, this behavior shows that the dynamical 
evolution becomes unstable, but all the physical quantities are finite.

\section{Late-time Behavior of the Singular Inflation Models}

The singular inflation models we presented in the previous sections can 
potentially have a quite interesting late-time behavior, from a 
phenomenological point of view. In this section we properly modify the models 
we worked out in the previous section, so that to achieve the unification of 
early-time and late-time acceleration with the same model. Note that assuming 
the prefect fluid form \cite{reviews1}, the effective equation of state (EoS), 
for a general bigravity model is given by,
\begin{equation}\label{eosdeded}
w_\mathrm{eff}=-1-\frac{2\dot{H}}{3H^2}\, ,
\end{equation}
As a first example that late-time and early-time acceleration occurs, we can 
think as follows: In the metric of Eq.~(\ref{AA5}), if the scale factor 
$a_J(t)$ behaves as $a_J(t)^2 = \frac{l^2}{t^2}$ or equivalently if the Hubble 
rate behaves as $H_J(t)= - \frac{1}{t}$, then the metric describes a de Sitter 
evolution of the Universe. Note that the parameter $l$ appearing in the scale 
factor $a_J(t)$ and in the Hubble rate $H_J(t)$, is a constant of length 
dimension. In the case that the scale factor behaves as $\tilde a(t)^2 = 
\frac{l^{2n}}{t^{2n}}$ or equivalently the Hubble rate behaves as $H_J(t) = - 
\frac{n}{t}$ with $n\neq 1$, if $0<n<1$, then the metric corresponds to a 
phantom evolution of the Universe. In addition, in the case that $n>1$, the 
Universe's evolution is a quintessential acceleration, and in the case that 
$n<0$, the Universe decelerating. We should note that in order for the Universe 
to be expanding, we should have $t<0$ when $n>0$, and in addition, $t>0$ when 
$n<0$. Then, if we consider Einstein gravity coupled with a perfect fluid with 
its EoS parameter being equal to,
\be
\label{AA19}
w = - \frac{1}{3} \left( 2 + \frac{1}{n} \right)\, ,
\ee
then we find that $n = - \frac{1}{3w + 2}$. Hence, in the case that the perfect 
fluid describes collisionless dust, we must have $w=0$ and therefore, $n$ must 
be $n=- \frac{1}{2}$.

We now turn our focus on another example with interesting early-time and 
late-time acceleration, for which the unification of these two accelerating 
eras can be achieved. We will assume that a Type IV is incorporated in the 
model, so the Hubble rate is equal to,
\be
\label{AA20}
H(t) = \left( H_0 + H_1 \left| t - t_s \right|^\gamma \right) \frac{\e^{- 
\frac{t-t_1}{l_1}}}{\e^{- \frac{t-t_1}{l_1}} + 1}
+ \frac{1}{ 2 \left( t_m + t \right) } \e^{ - \frac{ \left( t - t_2 \right)^2 
}{ l_2^2}}
  - \frac{1}{t} \frac{\e^{ \frac{t-t_3}{l_3}}}{\e^{- \frac{t-t_3}{l_3}} + 1} \, 
.
\ee
The model contains a lot of free parameters, so we impose the following 
restriction on these: $t_s < t_1 < t_2 < t_3 < 0 < t_m$ and $l_1$, $l_2$, 
$l_3>0$ and in addition we assume that $t$ is negative. The EoS for the Hubble 
rate of Eq.~(\ref{AA20}) is equal to,
\begin{align}\label{eos1}
w_{\mathrm{eff}}=&-1-\frac{2 
\left(-\frac{1}{\left(1+\e^{1-\frac{t}{t_3}}\right)^2 t t_3}+\frac{\e^{-1+\frac{2 
t}{t_3}} (-t+t_3)}{\left(\e+\e^{t/t_3}\right) t^2 
t_3}+\frac{\e^{-\frac{(t-t_2)^2}{t_2^2}} (-t+t_2)}{t_2^2 
(1+t_m)}\right)}{3\left(-\frac{\e^{-1+\frac{2 t}{t_3}}}{\e t+\ e^{t/t_3} 
t}+\frac{\e^{-\frac{(t-t_2)^2}{t_2^2}}}{2 (1+t_m)}+\frac{\e \left(H_0+H_1 
(t-t_s)^{\gamma }\right)}{\e+\e^{t/t_1}}\right)^2} \nn
& -\frac{2 \left(\frac{\e^2 \left(H_0+H_1 (t-t_s)^{\gamma 
}\right)}{\left(\e+\e^{t/t_1}\right)^2 t_1}+\frac{\e \left(H_0 (-t+t_s)-H_1 
(t-t_s)^{\gamma } (t-t_s-t_1 \gamma )\right)}{\left(\e+\e^{t/t_1}\right) t_1 
(t-t_s)}\right)}{3\left(-\frac{\e^{-1+\frac{2 t}{t_3}}}{\e t+\e^{t/t_3} 
t}+\frac{\e^{-\frac{(t-t_2)^2}{t_2^2}}}{2 (1+t_m)}+\frac{\e \left(H_0+H_1 
(t-t_s)^{\gamma }\right)}{\e+\e^{t/t_1}}\right)^2}\, .
\end{align}
Then, when $t<t_1$, the first term dominates and a Type IV singularity occurs 
at $t=t_s$. Near the singularity, when $t\simeq t_s$, the Hubble rate is 
approximately equal to,
\be
\label{AA20B}
H(t) \simeq \left( H_0 + H_1 \left| t - t_s \right|^\gamma \right) \frac{\e^{- 
\frac{t-t_1}{l_1}}}{\e^{- \frac{t-t_1}{l_1}} + 1} \, ,
\ee
while the effective EoS in this case is approximately equal to,
\begin{align}\label{eos2}
& w_{\mathrm{eff}}\simeq -1-\frac{2 
\left(-\frac{1}{\left(1+\e^{1-\frac{t}{t_3}}\right)^2 t t_3}+\frac{\e^{-1+\frac{2 
t}{t_3}} (-t+t_3)}{\left(\e+\e^{t/t_3}\right) t^2 
t_3}\right)}{3\left(-\frac{\e^{-1+\frac{2 t}{t_3}}}{\e t+\e^{t/t_3} t}+\frac{\e 
\left(H_0+H_1 (t-t_s)^{\gamma }\right)}{\e+\e^{t/t_1}}\right)^2}-\frac{2 
\left(\frac{\e^2 \left(H_0+H_1 (t-t_s)^{\gamma 
}\right)}{\left(\e+\e^{t/t_1}\right)^2 t_1}+\frac{\e \left(H_0 (-t+t_s)-H_1 
(t-t_s)^{\gamma } (t-t_s-t_1 \gamma )\right)}{\left(\e+\e^{t/t_1}\right) t_1 
(t-t_s)}\right)}{3\left(-\frac{\e^{-1+\frac{2 t}{t_3}}}{\e t+\e^{t/t_3} t}+\frac{\e 
\left(H_0+H_1 (t-t_s)^{\gamma }\right)}{\e+\e^{t/t_1}}\right)^2}\, ,
\end{align}
Consequently, near the Type IV singularity, the EoS can be further simplified 
to the following expression,
\begin{equation}\label{simeos1}
w_{\mathrm{eff}}\simeq -1-\frac{2 
\left(-\frac{1}{\left(1+\e^{1-\frac{t}{t_3}}\right)^2 t t_3}+\frac{\e^{-1+\frac{2 
t}{t_3}} (-t+t_3)}{\left(1+\e^{t/t_3}\right) t^2 
t_3}\right)}{3\left(-\frac{\e^{-1+\frac{2 t}{t_3}}}{ t+\e^{t/t_3} t}+\frac{ 
(H_0)}{\e+\e^{t/t_1}}\right)^2}-\frac{2 \left(\frac{\e^2 
(H_0)}{\left(\e+\e^{t/t_1}\right)^2 t_1}\right)}{3\left(-\frac{\e^{-1+\frac{2 
t}{t_3}}}{ t+\e^{t/t_3} t}+\e^{t/t_1}\right)^2}\, .
\end{equation}
The EoS of Eq.~(\ref{simeos1}) describes a quintessential acceleration if 
$t>t_3/2$, and phantom acceleration if otherwise. In the case that $t \sim 
t_2$, the second term dominates and a nearly de Sitter Universe can be 
realized, since the EoS in this case is approximately equal to,
\begin{equation}\label{simeos2}
w_{\mathrm{eff}}\simeq -1-\frac{8 \e^{\frac{(t-t_2)^2}{t_2^2}} (-t+t_2) 
(1+t_m)}{3t_2^2}\, ,
\end{equation}
and since $t\simeq t_2$, we get $w_{\mathrm{eff}}\simeq -1$. Furthermore when 
$t>t_3$, the last term dominates in Eq.~(\ref{AA20B}) and the Universe becomes 
again a quintessential accelerating Universe, since the EoS in this case is,
\begin{equation}\label{simeos3}
w_{\mathrm{eff}}\simeq \frac{1}{3} \left(-3+2 \e^{2-\frac{4 t}{t_3}} \left(\e 
t+\e^{t/t_3} t\right)^2 \left(\frac{1}{t t_3}+\frac{\e^{-1+\frac{2 t}{t_3}} 
(t-t_3)}{\left(\e+\e^{t/t_3}\right) t^2 t_3}\right)\right)\, .
\end{equation}

As a final quite phenomenologically interesting model, we shall consider a 
modified variant singular form of the $R^2$ inflation model of 
Eq.~(\ref{singstarobhub}), for which the Hubble rate reads,
\begin{equation}\label{hubblemodform}
H(t)=\frac{2}{3 \left(\frac{4}{3 H_0}+t\right)}+\e^{-(t-t_s)^{\gamma }} 
\left(\frac{H_0}{2}+H_i (t-t_i)\right)+f_0 (t-t_0)^{\delta } (t-t_s)^{\gamma 
}\, .
\end{equation}
where $\delta$ and $f_0$ are arbitrary constant and positive parameters. In 
this case we assume that $\gamma,\delta>1$, so that the cosmological evolution 
develops two Type IV singularities, one at $t=t_s$ and one at $t=t_0$. Also the 
cosmological time $t_s$ is assumed to be at the end of inflation and the time 
instance $t_0$ is assumed to be much more later than $t_s$, so that $t_s\ll 
t_0$. Also if $t_p$ represents the present time, then $t_0\ll t_p$, so 
practically $t_0$ characterizes an intermediate cosmological era of evolution. 
In addition, the parameters $H_0,H_i$ which are related to the Starobinsky 
model, are constrained by observational data to satisfy $H_0,H_i\gg 1$ (see 
\cite{noo4}). Before we proceed, let us see which bigravity model can generate 
the cosmological evolution of Eq.~(\ref{hubblemodform}), and the bigravity can 
be determined by calculating the functions $\omega (t)$, $V(t)$, $\sigma (t)$ 
and $U(t)$ appearing in Eq.~(\ref{Fbi19D}). These functions for the Hubble rate 
of Eq.~(\ref{Fbi19D}) read,
\begin{align}\label{functionsforbigravity}
\omega (t)=& 2 \sqrt{3} \left(\frac{2}{3 \left(\frac{4}{3 
H_0}+t\right)}+\e^{-(t-t_s)^{\gamma }} \left(\frac{H_0}{2}+H_i 
(t-t_i)\right)+f_0 (t-t_0)^{\delta } (t-t_s)^{\gamma }\right) \, , \nn
\tilde V(t)=&-6 \left(\frac{2}{3 \left(\frac{4}{3 
H_0}+t\right)}+\e^{-(t-t_s)^{\gamma }} \left(\frac{H_0}{2}+H_i 
(t-t_i)\right)+f_0 (t-t_0)^{\delta } (t-t_s)^{\gamma }\right)^2\, , \nn
\sigma (t)= & \frac{12 M_g^2 \left(\frac{2}{3 \left(\frac{4}{3 
H_0}+t\right)}+\e^{-(t-t_s)^{\gamma }} \left(\frac{H_0}{2}+H_i 
(t-t_i)\right)+f_0 (t-t_0)^{\delta } (t-t_s)^{\gamma }\right)^2}{M_f^2}\, , \nn
\tilde U(t)= & -6 \frac{M_g^2}{M_f^2} \left(m^2 M_\mathrm{eff}^2+6 \left(\frac{2}{3 
\left(\frac{4}{3 H_0}+t\right)}+\e^{-(t-t_s)^{\gamma }} \left(\frac{H_0}{2}+H_i 
(t-t_i)\right)+f_0 (t-t_0)^{\delta } (t-t_s)^{\gamma }\right)^2\right) \nn
& \times \left(\frac{2}{3 \left(\frac{4}{3 
H_0}+t\right)}+\e^{-(t-t_s)^{\gamma }} \left(\frac{H_0}{2}+H_i 
(t-t_i)\right)+f_0 (t-t_0)^{\delta } (t-t_s)^{\gamma }\right)^2\, ,
\end{align}
and again we took into account Eq.~(\ref{AA6}), so the above relations hold 
true near the singularity, where the conformal time is nearly identical to the 
cosmic time. Let us now proceed to the phenomenological implications of the 
model of Eq.~(\ref{hubblemodform}). As we demonstrate, it has quite interesting 
phenomenological implications, since the early-time acceleration, the matter 
domination and also the late-time acceleration eras can be described by a 
single model. This is already obvious from the functional form of the Hubble 
rate, since at $t\simeq t_s$,  the Hubble rate becomes approximately equal to,
\begin{equation}\label{approx}
H(t)\simeq H_0+H_i(t-t_i)\, ,
\end{equation}
since the term $\sim \frac{2}{3 \left(\frac{4}{3 H_0}+t\right)}$ is 
approximately equal to $\sim \frac{H_0}{2}$ and also the term $\sim 
(t-t_s)^{\gamma}$ is approximately equal to zero. So near $t=t_s$, which is 
assumed to occur near the early-time acceleration era, the model becomes nearly 
the Starobinsky $R^2$ inflation model, which is in concordance with 
observations \cite{planck,noo4}. Let us now investigate what happens as the 
cosmic time evolves, and the best way to study the phenomenology is to study 
the EoS. The general form of the EoS for the Hubble rate (\ref{hubblemodform}), 
is equal to,
\begin{align}
\label{generaleosfor modstarmod}
w_{\mathrm{eff}}=&-1-\frac{2 \left(\e^{-(t-t_s)^{\gamma }} H_i-\frac{2}{3 
\left(\frac{4}{3 H_0}+t\right)^2}+f_0 (t-t_0)^{\delta } (t-t_s)^{-1+\gamma } 
\gamma \right)}{3 \left(\frac{2}{3 \left(\frac{4}{3 
H_0}+t\right)}+\e^{-(t-t_s)^{\gamma }} \left(\frac{H_0}{2}+H_i 
(t-t_i)\right)+f_0 (t-t_0)^{\delta } (t-t_s)^{\gamma }\right)^2} \nn
& -\frac{2 \left(-\e^{-(t-t_s)^{\gamma }} \left(\frac{H_0}{2}+H_i (t-t_i)\right) 
(t-t_s)^{-1+\gamma } \gamma +f_0 (t-t_0)^{-1+\delta } (t-t_s)^{\gamma } \delta 
\right)}{3 \left(\frac{2}{3 \left(\frac{4}{3 H_0}+t\right)}+\e^{-(t-t_s)^{\gamma 
}} \left(\frac{H_0}{2}+H_i (t-t_i)\right)+f_0 (t-t_0)^{\delta } (t-t_s)^{\gamma 
}\right)^2}
\, ,
\end{align}
so at early times and near $t\simeq t_s$, the EoS becomes,
\begin{equation}\label{eosearlyts}
w_{\mathrm{eff}}\simeq -1-\frac{2 \left(\frac{3 H_0}{4}+H_i\right)}{3 (H_0+H_i 
(t-t_i))^2}\, ,
\end{equation}
and since a viable $R^2$ inflation model requires that $H_0$ and $H_i$ to be 
very large \cite{noo4}, the EoS is approximately equal to 
$w_{\mathrm{eff}}\simeq -1$, which is a de Sitter accelerating phase, as was 
expected. At $t\simeq t_0$, which is much after the time instance $t_s$, the 
EoS becomes approximately equal to,
\begin{equation}\label{eosearlyts1}
w_{\mathrm{eff}}\simeq -1-\frac{2 \left(\e^{-(t-t_s)^{\gamma }} H_i-\frac{2}{3 
\left(\frac{4}{3 H_0}+t\right)^2}-\e^{-(t-t_s)^{\gamma }} 
\left(\frac{H_0}{2}+H_i (t-t_i)\right) (t-t_s)^{-1+\gamma } \gamma \right)}{3 
\left(\frac{2}{3 \left(\frac{4}{3 H_0}+t\right)}+\e^{-(t-t_s)^{\gamma }} 
\left(\frac{H_0}{2}+H_i (t-t_i)\right)\right)^2}\, .
\end{equation}
It is obvious that since $t\sim t_0\gg t_s$, and also $t\gg \frac{4}{3 H_0}$ 
the above expression becomes,
\begin{equation}\label{eosearlyts1a}
w_{\mathrm{eff}}\simeq -1-\frac{2 \left(\e^{-t^{\gamma }} H_i-\frac{2}{3 
t^2}-\e^{-t^{\gamma }} \left(\frac{H_0}{2}+H_i (t-t_i)\right) t^{-1+\gamma } 
\gamma \right)}{3 \left(\frac{2}{3 t}+\e^{-t^{\gamma }} \left(\frac{H_0}{2}+H_i 
(t-t_i)\right)\right)^2}\, ,
\end{equation}
and since the $t\gg 1$, the terms proportional to $\sim \e^{-t^{\gamma }}$ are 
exponentially suppressed. In effect, the EoS becomes,
\begin{equation}\label{eosearlyts2}
w_{\mathrm{eff}}\simeq -1-\frac{2 \left(-\frac{2}{3 t^2}\right)}{3 
\left(\frac{2}{3 t}\right)^2}\, ,
\end{equation}
which is approximately equal to zero. So the era near $t\simeq t_0$ describes a 
matter domination era in this case since $w_{\mathrm{eff}}\simeq 0$. Finally, 
at $t\simeq t_p$, the EoS becomes approximately equal to,
\begin{equation}\label{eosearlyts3}
w_{\mathrm{eff}}\simeq -1-\frac{2 \left(\e^{-t^{\gamma }} H_i-\frac{2}{3 
t^2}+f_0 t^{-1+\gamma +\delta } \gamma - \e^{-t^{\gamma }} t^{-1+\gamma } 
\left(\frac{H_0}{2}+H_i (t-t_i)\right) \gamma +f_0 t^{-1+\gamma +\delta } 
\delta \right)}{3 \left(\frac{2}{3 t}+f_0 t^{\gamma +\delta }+ \e^{-t^{\gamma }} 
\left(\frac{H_0}{2}+H_i (t-t_i)\right)\right)^2}\, ,
\end{equation}
so by omitting the exponentially suppressed terms and also since the term $\sim 
\frac{1}{t}$ is subdominant at times $t\sim t_p$, compared to the positive 
powers of $t$, the EoS becomes finally,
\begin{equation}\label{eosearlyts4}
w_{\mathrm{eff}}\simeq -1-\frac{2 t^{-1-\gamma -\delta } \gamma }{3 
f_0}-\frac{2 t^{-1-\gamma -\delta } \delta }{3 f_0}\, .
\end{equation}
Since $t\sim t_p$ and $t_p$ is approximately of the order $t\sim 10^{17}$sec, 
this means that the terms $\sim t^{-1-\gamma -\delta }$ satisfy $t^{-1-\gamma 
-\delta}\ll 1$, owing to the fact that we initially assumed $\gamma, \delta 
>1$, so that two Type IV singularities occur. Therefore, at late-time we have 
$w_{\mathrm{eff}}\simeq -1$, and hence, the Universe is described by a nearly 
de Sitter evolution, slightly crossing the phantom divide. In fact, such an 
evolution is supported by present time observations which predict that 
$w_{\mathrm{eff}}\simeq -1$, but with the EoS slightly crossing the phantom 
divide \cite{phantom}. Hence, with the model of (\ref{hubblemodform}) we were 
able to describe within the same theoretical framework, three cosmological 
eras, an early de Sitter acceleration era, a matter domination era, and a 
late-time acceleration era. In Table~\ref{phenstarob} we present the behavior 
of the EoS for the modified singular $R^2$ inflation model of 
Eq.~(\ref{hubblemodform}) for the various cosmological eras.


\begin{table*}[h]
\small
\caption{\label{phenstarob}Behavior of Equation of State for the Modified 
Singular $R^2$ Inflation Model of Eq.~(\ref{hubblemodform})}
\begin{tabular}{@{}ccccc@{}}
\tableline
\tableline
\tableline
Cosmological Time & \ \qquad \  & EoS $w_{\mathrm{eff}}$ & \ \qquad \  &Evolution Type \\
\tableline
$t \simeq t_s$ && $w_{\mathrm{eff}}\simeq -1$ && Nearly de Sitter \\
\tableline
$t \simeq t_0$ && $w_{\mathrm{eff}}\simeq 0$ && Matter Domination \\
\tableline
$t \simeq t_p$ && $w_{\mathrm{eff}}\simeq -1$ && Nearly de Sitter \\
\tableline
\tableline
  \end{tabular}
\end{table*}

\section{Conclusions}

In this paper we demonstrated how a Type IV singular evolution can be 
successfully generated by a bimetric $F(R)$ gravity theory. Particularly, by 
employing the formalism of bimetric $F(R)$ gravity theory, we were able to 
describe two singular evolutions, one of which is a singular variant of the 
Starobinsky $R^2$ inflation model. By calculating the slow-roll parameters 
$\epsilon$, $\eta$ and $\xi$, we showed that in some cases these parameters 
contain singularities. Hence, one could claim that the observable quantities 
are singular and therefore the singularities lead to unphysical results, but 
this is not the case, since the singularities in the slow-roll parameters 
indicate that a strong instabilities occur at the time instance that the 
singularity occurs. Indeed, the slow-roll parameters are the lowest order terms 
in the slow-roll expansion, and hence an abrupt singular increase of their 
values indicates that the slow-roll expansion breaks down. Therefore, the 
cosmological dynamical system becomes unstable at the singularity point and 
therefore the solution that described the inflationary solution up to that 
point, ceases to be the final attractor of the theory and therefore a new 
attractor is chosen by the theory. Hence, the singularities in the slow-roll 
parameters show that graceful exit is triggered at that point. Of course the 
presence of singularities per se is not sufficient for proving that inflation 
indeed ends at the singularity, but another underlying mechanism probably 
controls the exit, like curvature perturbations around unstable de Sitter 
vacua, as it was the case in \cite{sergeitraceanomaly,noo4}, or a tachyonic 
instability exists in the theory.

Apart from the early-time behavior, we examined the late-time behavior of the 
models we mentioned earlier, and we showed that the unified description of 
early and late-time acceleration was possible in the context of bimetric $F(R)$ 
gravity. In addition, in one of the models we studied, we showed that three 
different eras can be successfully described by using a single model, namely, 
early time acceleration, the matter domination era, and also the late-time 
acceleration era.

What would be of fundamental importance is to find a mechanism to describe the 
graceful exit from inflation, since in the present study we provided some 
sufficient proof that graceful exit might occur at the time instance that the 
singularity occurs, but we did not proved that graceful exit indeed occurs. It is 
therefore important to investigate in the Jordan frame, with which mechanism 
the graceful exit can actually occur. In addition, we should also investigate 
if the second metric, the fiducial one, can be distinguished phenomenologically 
from the physical metric, at the level of cosmological perturbations. In this 
paper we took into account a flat Minkowski fiducial metric, but this is the 
simplest case one can choose. In principle, one could also choose a non-flat 
FRW metric, so this could perplex the study to a great extent. This 
question will be considered elsewhere.

\section*{Acknowledgments}

This work is supported by MINECO (Spain), project
  FIS2013-44881 and I-LINK 1019 (S.D.O) and by Min. of Education and 
Science of Russia 
(S.D.O
and V.K.O) and  (in part) by
MEXT KAKENHI Grant-in-Aid for Scientific Research on Innovative Areas ``Cosmic
Acceleration''  (No. 15H05890) and the JSPS Grant-in-Aid for Scientific 
Research (C) \# 23540296 (S.N.).

\section*{Appendix: Detailed Form of Some Intermediate Expressions}

In this Appendix we present the detailed form of the slow-roll indices 
corresponding to the singular $R^2$ model with Hubble rate appearing in 
Eq.~(\ref{singstarobhub}). We start off with the parameter $\epsilon$, with its 
full form being equal to,
\begin{align}
\label{sgfdfd}
\epsilon\simeq& -3 \left( \frac{\left(H_i-\frac{1}{6} M^2 (t-t_i)+f_0 
(t-t_s)^{\gamma }\right)^2 \left(-6 H_i+M^2 (-1+t-t_i)-6 f_0 (t-t_s)^{-1+\gamma 
} (t-t_s-\gamma )\right)}{Q(t)} \right. \nn
& +\frac{+\frac{1}{36} \left(6 H_i-M^2 (-1+t-t_i)+6 f_0 (t-t_s)^{-1+\gamma } 
(t-t_s-\gamma )\right)^2+\frac{M^2\left(H_i-\frac{1}{6} M^2 (t-t_i)+f_0 
(t-t_s)^{\gamma }\right)^2}{3}}{Q(t)} \nn
& +\frac{\left(H_i-\frac{1}{6} M^2 (t-t_i)+f_0 (t-t_s)^{\gamma }\right)^4-2 f_0 
(t-t_s)^{-1+\gamma } \gamma \left(H_i-\frac{1}{6} M^2 (t-t_i)+f_0 
(t-t_s)^{\gamma }\right)^2}{Q(t)} \nn
& \left. + \frac{f_0 (t-t_s)^{-2+\gamma } (-1+\gamma ) \gamma -\left(-\frac{M^2}{6}+f_0 
(t-t_s)^{-1+\gamma } \gamma \right)^2}{Q(t)}
\right)\, ,
\end{align}
where the function $Q(t)$ stands for,
\begin{align}
\label{sgfdfd2}
Q(t) =&2 \left(H_i-\frac{1}{6} M^2 (t-t_i)+f_0 (t-t_s)^{\gamma }\right)^5 
\left(3+\frac{-6 H_i+M^2 (-1+t-t_i)-6 f_0 (t-t_s)^{-1+\gamma } (t-t_s-\gamma 
)}{6 \left(H_i-\frac{1}{6} M^2 (t-t_i)+f_0 (t-t_s)^{\gamma }\right)^3}\right)^2 
\nn 
& \times \left(-6 H_i+M^2 (-1+t-t_i)-6 f_0 (t-t_s)^{-1+\gamma } 
(t-t_s-\gamma )\right)\, .
\end{align}
Correspondingly, the full form of the slow-roll parameter $\eta$ is,
\begin{align}
\label{etafullform}
\eta\simeq & -\frac{1}{4 \left(3-\frac{36 \left(6 H_i+M^2 
(1-t+t_i)\right)}{\left(6 H_i+M^2 (-t+t_i)\right)^3}\right)} \nn 
& \times \left( -\frac{648 \left(6 H_i+M^2 (1-t+t_i)\right)}{\left(6 H_i+M^2 
(-t+t_i)\right)^3}+\frac{36 \left(6 H_i+M^2 (1-t+t_i)\right)^2}{\left(6 H_i+M^2 
(-t+t_i)\right)^4} \right. \nn
& +\frac{36 
\left(\frac{M^2}{3}+\left(H_i+\frac{1}{6} M^2 
(-t+t_i)\right)^2-\frac{M^4}{\left(6 H_i+M^2 (-t+t_i)\right)^2}+\frac{f_0 
(t-t_s)^{-2+\gamma } (-1+\gamma ) \gamma }{\left(H_i+\frac{1}{6} M^2 
(-t+t_i)\right)^2}\right)}{-6 H_i+M^2 (-1+t-t_i)} \nn 
& +\frac{3 \left(\frac{M^2}{3}+\left(H_i+\frac{1}{6} M^2 
(-t+t_i)\right)^2-\frac{M^4}{\left(6 H_i+M^2 (-t+t_i)\right)^2}+\frac{f_0 
(t-t_s)^{-2+\gamma } (-1+\gamma ) \gamma }{\left(H_i+\frac{1}{6} M^2 
(-t+t_i)\right)^2}\right)}{\left(H_i+\frac{1}{6} M^2 (-t+t_i)\right)^2} \nn
&-\frac{\left(\frac{M^2}{3}+\left(H_i+\frac{1}{6} M^2 
(-t+t_i)\right)^2-\frac{M^4}{\left(6 H_i+M^2 (-t+t_i)\right)^2}+\frac{f_0 
(t-t_s)^{-2+\gamma } (-1+\gamma ) \gamma }{\left(H_i+\frac{1}{6} M^2 
(-t+t_i)\right)^2}\right)^2}{\left(H_i+\frac{1}{6} M^2 (-t+t_i)\right)^4} \nn
& + \frac{12 \left(-H_i^2+\frac{1}{3} H_i M^2 (t-t_i)+\frac{1}{36} M^4 
\left(-t^2+2 t t_i-t_i^2-\frac{36}{\left(6 H_i+M^2 
(-t+t_i)\right)^2}\right)\right)}{-6 H_i+M^2 (-1+t-t_i)} \nn
& \left. + \frac{12 \left(-\frac{36 f_0 (t-t_s)^{-3+\gamma } (2+3 t-3 t_s-\gamma ) 
(-1+\gamma ) \gamma }{\left(6 H_i+M^2 (-t+t_i)\right)^2}+M^2 
\left(-\frac{5}{6}+\frac{72 f_0 (t-t_s)^{-2+\gamma } (-1+\gamma ) \gamma 
}{\left(6 H_i+M^2 (-t+t_i)\right)^3}\right)\right)}{-6 H_i+M^2 
(-1+t-t_i)} \right) \, .
\end{align}
Finally, the full form of slow-roll parameter $\xi$ is too lengthy to quote 
here, so we give only the terms that contain singularities, so the singular 
part of $\xi$, which we denote as $\xi_s$, is equal to,
\begin{align}
\label{singularityxi}
\xi_s= & \frac{f_0 (t-t_s)^{-2+\gamma } (-1+\gamma ) \gamma  }{4 
\left(H_i+\frac{1}{6} M^2 (-t+t_i)\right)^4 \left(3-\frac{36 \left(6 H_i+M^2 
(1-t+t_i)\right)}{\left(6 H_i+M^2 (-t+t_i)\right)^3}\right)} \nn 
& \times \left( \frac{4 f_0 (t-t_s)^{-2+\gamma } (-1+\gamma ) \gamma 
}{\left(H_i+\frac{1}{6} M^2 (-t+t_i)\right)^4}+\frac{90 f_0 (t-t_s)^{-2+\gamma 
} (-1+\gamma ) \gamma }{\left(-6 H_i+M^2 (-1+t-t_i)\right) 
\left(H_i+\frac{1}{6} M^2 (-t+t_i)\right)^2} \right. \nn
& -\frac{648 f_0^2 
(t-t_s)^{-4+2 \gamma } (-1+\gamma )^2 \gamma ^2}{\left(-6 H_i+M^2 
(-1+t-t_i)\right)^3 \left(H_i+\frac{1}{6} M^2 (-t+t_i)\right)^2}-\frac{36 f_0^2 
(t-t_s)^{-4+2 \gamma } (-1+\gamma )^2 \gamma ^2}{\left(H_i+\frac{1}{6} M^2 
(-t+t_i)\right)^4 \left(6 H_i+M^2 (1-t+t_i)\right)^2} \nn 
& +\frac{1296 f_0^3 (t-t_s)^{-6+3 \gamma } (-1+\gamma )^3 \gamma 
^3}{\left(H_i+\frac{1}{6} M^2 (-t+t_i)\right)^4 \left(6 H_i+M^2 
(1-t+t_i)\right)^4} \nn
& +\frac{30 \left(\frac{72 f_0 M^2 (t-t_s)^{-2+\gamma } 
(-1+\gamma ) \gamma }{\left(6 H_i+M^2 (-t+t_i)\right)^3}-\frac{36 f_0 
(t-t_s)^{-3+\gamma } (2+3 t-3 t_s-\gamma ) (-1+\gamma ) \gamma }{\left(6 
H_i+M^2 (-t+t_i)\right)^2}\right)}{-6 H_i+M^2 (-1+t-t_i)} \nn 
& +\frac{3 
M^4 \left(H_i+\frac{1}{6} M^2 (-t+t_i)\right)^2 \left(\frac{72 f_0 M^2 
(t-t_s)^{-2+\gamma } (-1+\gamma ) \gamma }{\left(6 H_i+M^2 
(-t+t_i)\right)^3}-\frac{36 f_0 (t-t_s)^{-3+\gamma } (2+3 t-3 t_s-\gamma ) 
(-1+\gamma ) \gamma }{\left(6 H_i+M^2 (-t+t_i)\right)^2}\right)}{\left(6 
H_i+M^2 (1-t+t_i)\right)^2} \nn
&-\frac{432 f_0 (t-t_s)^{-2+\gamma } 
(-1+\gamma ) \gamma  \left(\frac{72 f_0 M^2 (t-t_s)^{-2+\gamma } (-1+\gamma ) 
\gamma }{\left(6 H_i+M^2 (-t+t_i)\right)^3}-\frac{36 f_0 (t-t_s)^{-3+\gamma } 
(2+3 t-3 t_s-\gamma ) (-1+\gamma ) \gamma }{\left(6 H_i+M^2 
(-t+t_i)\right)^2}\right)}{\left(-6 H_i+M^2 (-1+t-t_i)\right)^3} \nn 
& +\frac{36 \left(H_i+\frac{1}{6} M^2 (-t+t_i)\right)^2 }{\left(6 H_i+M^2 
(1-t+t_i)\right)^2} \left(-\frac{4 f_0 (t-t_s)^{-3+\gamma } (-2+\gamma ) 
(-1+\gamma ) \gamma }{\left(H_i+\frac{1}{6} M^2 (-t+t_i)\right)^2} \right. \nn
& \left. +\frac{f_0 
(t-t_s)^{-4+\gamma } (-3+\gamma ) (-2+\gamma ) (-1+\gamma ) \gamma 
}{\left(H_i+\frac{1}{6} M^2 (-t+t_i)\right)^2}\right) 
+\frac{f_0 
(t-t_s)^{-2+\gamma } (-1+\gamma ) \gamma  }{4 \left(H_i+\frac{1}{6} M^2 
(-t+t_i)\right)^4 \left(3-\frac{36 \left(6 H_i+M^2 (1-t+t_i)\right)}{\left(6 
H_i+M^2 (-t+t_i)\right)^3}\right)} \nn 
& \times \left( \frac{4 f_0 
(t-t_s)^{-2+\gamma } (-1+\gamma ) \gamma }{\left(H_i+\frac{1}{6} M^2 
(-t+t_i)\right)^4}+\frac{90 f_0 (t-t_s)^{-2+\gamma } (-1+\gamma ) \gamma 
}{\left(-6 H_i+M^2 (-1+t-t_i)\right) \left(H_i+\frac{1}{6} M^2 
(-t+t_i)\right)^2} \right. \nn 
& \left. \left. -\frac{648 f_0^2 (t-t_s)^{-4+2 \gamma } (-1+\gamma 
)^2 \gamma ^2}{\left(-6 H_i+M^2 (-1+t-t_i)\right)^3 \left(H_i+\frac{1}{6} M^2 
(-t+t_i)\right)^2}-\frac{36 f_0^2 (t-t_s)^{-4+2 \gamma } (-1+\gamma )^2 \gamma 
^2}{\left(H_i+\frac{1}{6} M^2 (-t+t_i)\right)^4 \left(6 H_i+M^2 
(1-t+t_i)\right)^2} \right) \right) \, .
\end{align}

\end{document}